\documentclass{proc}

\usepackage[preprint]{procjac}

\usepackage{color}

\docident{\hspace{\fill}
\textit{Presented at the Eleventh Workshop on RF Superconductivity}
\hspace{\fill}\makebox[0pt][r]{MSUCL-1276}}

\usepackage{graphicx}

\newcommand{\incgraphLANDHALF}[2]{%
\includegraphics[width=#2\columnwidth,%
bbllx=0.45in,bburx=5.15in,bblly=0.4in,bbury=4.2in]{#1}
}

\newcommand{\incgraphLANDHALFlab}[3]{%
\setlength{\unitlength}{0.69149in} 
\setlength{\unitlength}{#2\unitlength}
\begin{picture}(4.7,3.8)(-0.55,-0.6)
\put(-0.55,-0.6){\includegraphics[width=#2\columnwidth,%
bbllx=0.45in,bburx=5.15in,bblly=0.4in,bbury=4.2in]{#1}
}
\put(0.18,2.60){\Large #3}
\end{picture}
}

\newcommand{\degree}{\ensuremath{^\circ}}

\newcommand{\etal}{{\em et al.}}

\usepackage{hyperref}
\usepackage[all]{hypcap}

\hypersetup{%
pdftitle={Nb QWR Development for RIA},
pdfauthor={W. Hartung et al.}%
}

\begin{document}

\title{Niobium Quarter-Wave Resonator Development for the Rare Isotope
Accelerator\thanks{Work supported by the U.S. Department of Energy
under Grant DE-FG02-03ER41248.}}

\author{%
W.~Hartung, J.~Bierwagen, S.~Bricker, J.~Colthorp, C.~Compton, T.~Grimm,
S.~Hitchcock, F.~Marti,\\
L.~Saxton, R.~C.~York\\
National Superconducting Cyclotron Lab,
Michigan State University, East Lansing, Michigan\\
\and
A.~Facco, V.~Zviagintsev\\
Instituto Nazionale di Fisica Nucleare, Laboratori Nazionali di Legnaro,
Legnaro, Italy}

\maketitle


\section{Introduction}

There is a growing interest in the production of intense beams of exotic
isotopes for research in nuclear physics and laboratory astrophysics
\cite{R_JY}. The Rare Isotope Accelerator (RIA) is one such project being
pursued by the nuclear physics community in the USA \cite{R_JP}.  RIA calls for
a superconducting cavity linac to accelerate the CW beam of heavy ions to $\geq
400$ MeV per nucleon, with a beam power of up to 400 kW\@.  Several types of
superconducting structures are needed due to the changing velocity of the beam
\cite{R_IN}.

Design studies are in progress at Michigan State University (MSU) for a
10th-harmonic driver linac consisting of quarter-wave resonators (QWRs),
half-wave resonators, and elliptical cavities \cite{R_JZ}.  Three different
types of QWRs have been designed for the first segment of the driver linac. 
The first QWR type (optimum $\beta \equiv \beta_m = 0.041$, 80.5 MHz) is very
similar to existing QWRs in use at INFN-Legnaro ($\beta$ is the beam velocity
divided by the speed of light).  The second ($\beta_m = 0.085$, 80.5 MHz) and
third ($\beta_m = 0.16$, 161 MHz) types are being developed as a collaborative
effort between Legnaro and MSU.

This paper covers the RF design, prototyping, and preliminary RF testing of
simplified versions of the $\beta_m = 0.085$ and $\beta_m = 0.16$ QWRs.  The
next step, the development of a complete $\beta_m = 0.16$ cavity with an
integrated helium vessel, is also underway \cite{R_KA}.

\section{Cavity Design}

The quarter-wave resonators developed by Legnaro for ALPI are the basis for the
design of the RIA quarter-wave cavities.  QWRs at 80 MHz are presently being
used at Legnaro for the ALPI and PIAVE linacs \cite{R_KB}; a 160 MHz QWR has
also been prototyped at Legnaro \cite{R_KC}.

The ALPI cavities have a outer conductor diameter of 180 mm; this was enlarged
to 240 mm for the $\beta_m = 0.085$ and $\beta_m = 0.16$ RIA cavities.  A
larger aperture (30 mm) is also used for the RIA cavities.  Another new feature
is to separate the cavity vacuum from the insulation vacuum to reduce
particulate contamination of the cavity surfaces.

A recent development is the identification of beam steering due to the vertical
asymmetry in QWR structures \cite{R_KD}.  The steering is not a major problem
for the $\beta_m = 0.085$ QWR due to the long wavelength, but it is significant
in the $\beta_m = 0.16$ QWR\@.  It has been shown theoretically that the
steering can be partially compensated by asymmetric shaping of the cavity in
the vicinity of the beam ports \cite{R_KE}.  The $\beta_m = 0.16$ QWR 
incorporates asymmetric cavity walls to compensate for the steering.  Beam
dynamics studies indicate that the compensation should eliminate the emittance
growth due to steering \cite{R_KF}.

\begin{figure}
\begin{center}
\setlength{\unitlength}{0.01in}
\begin{picture}(325,364)
\put(31,0){\includegraphics[width=0.81\columnwidth,%
bbllx=255pt,bblly=88pt,bburx=565pt,bbury=516pt]{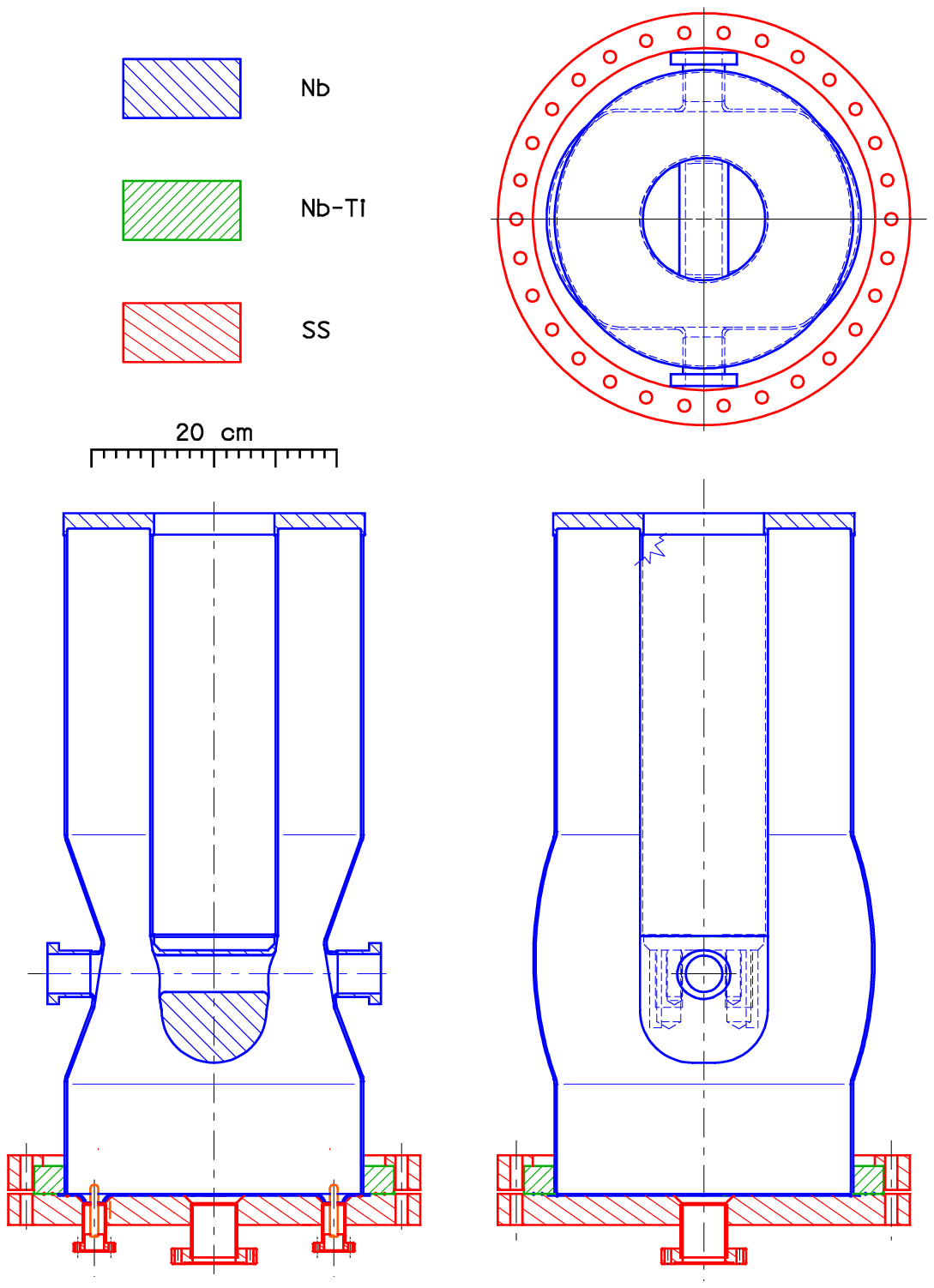}}
\put(0,289){\makebox(50,75){\Large(a)}}
\end{picture}\\[1ex]
\setlength{\unitlength}{0.01in}
\begin{picture}(325,524)
\put(31,0){\includegraphics[width=0.81\columnwidth,%
bbllx=464pt,bblly=62pt,bburx=822pt,bbury=774pt]{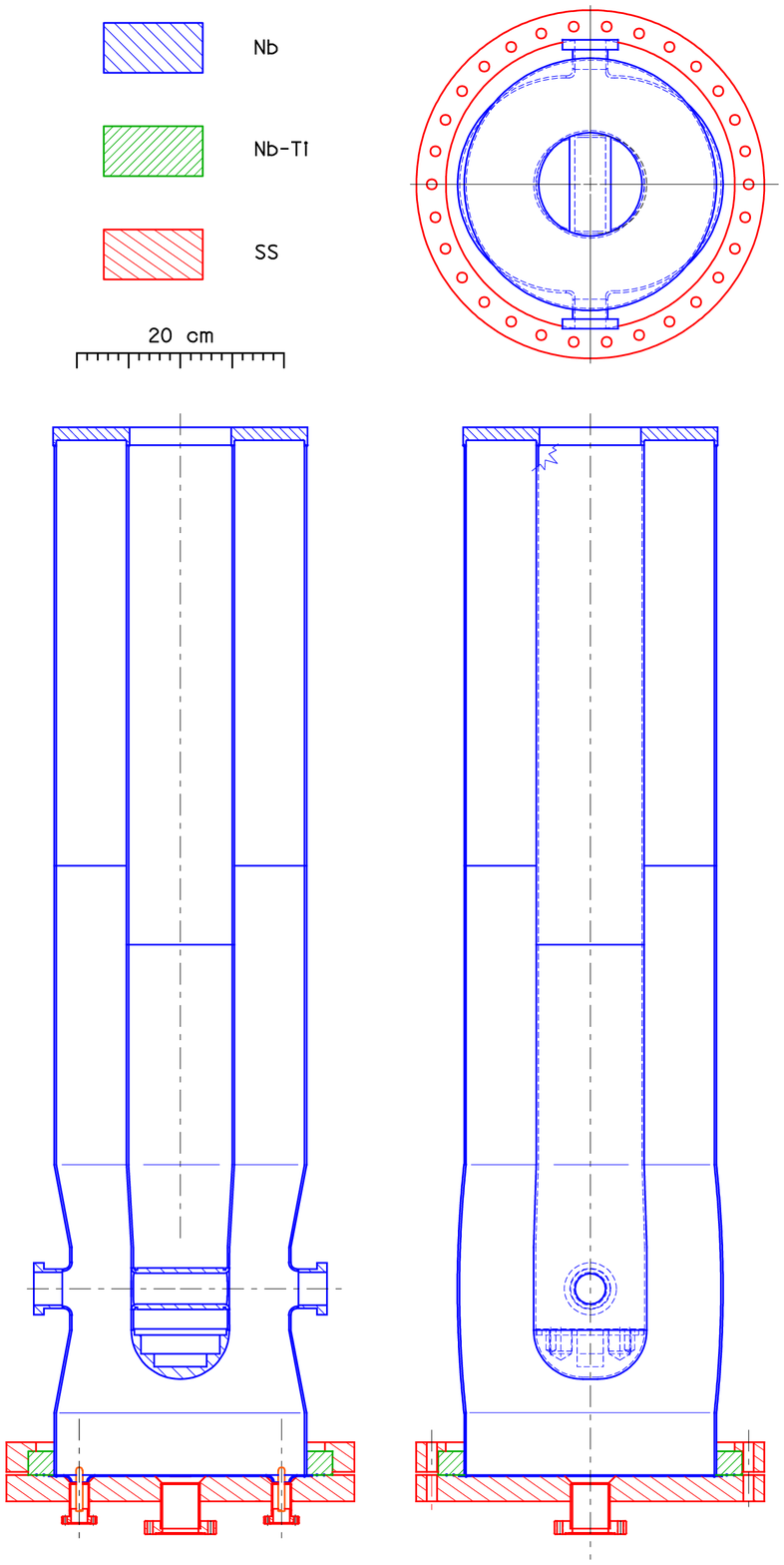}}
\put(0,449){\makebox(50,75){\Large(b)}}
\end{picture}\\
\end{center}

\caption{Three-view drawing of (a) the $\beta_m = 0.16$ QWR and (b) the
$\beta_m = 0.085$ QWR.\label{F:dwg}}
\end{figure}

\begin{figure}[tb]
\incgraphLANDHALF{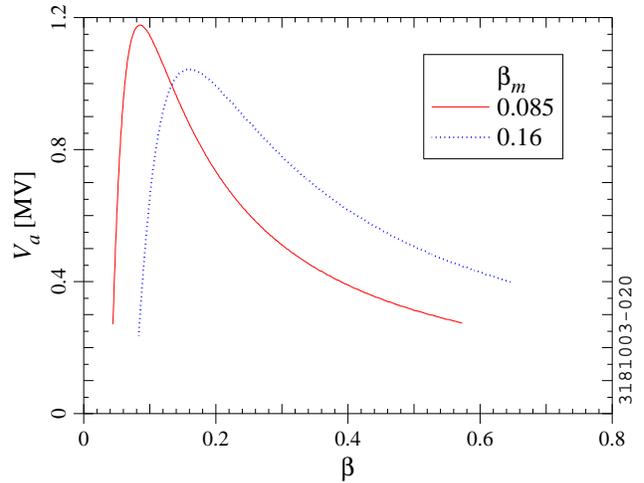}{1.0}

\caption{Dependence of accelerating voltage on beam velocity for the $\beta_m =
0.16$ and $\beta_m = 0.085$ QWRs.\label{F:TT}}
\end{figure}

\autoref{T:parms} shows some of the parameters of the structures.  RF
parameters were calculated with ANALYST.\footnote{A product of Simulation
Technology \& Applied Research, Inc., Mequon, Wisconsin.}  Drawings of the
structures are shown in \autoref{F:dwg}.  \autoref{F:TT} shows the
accelerating voltage that can be delivered by each of the cavity types at the
design field level ($E_p = 20$ MV/m) as a function of the velocity of the
accelerated beam, including transit time effects.

\begin{table}[tbh]
\caption{Selected RF and geometrical parameters for the quarter-wave
resonators.  OC and IC are the outer conductor and inner conductor,
respectively.\label{T:parms}}

\vspace*{1ex}

\begin{center}
\begin{tabular}{|l|c|c|}
\hline
Optimum $\beta \equiv \beta_m$ & 0.085 & 0.16 \\ \hline
Resonant frequency $f$ & 80.5 MHz & 161 MHz\\ \hline
Design $E_p$ & \multicolumn{2}{c|}{20 MV/m} \\ \hline
Design $B_p$ & 49.2 mT & 43.4 mT \\ \hline
Design $V_a$ & 1.18 MV & 0.99 MV \\ \hline
$R_s/Q$ & 416 $\Omega$ & 380 $\Omega$ \\ \hline
Geometry factor & 19 $\Omega$ & 35 $\Omega$ \\ \hline
\hline
Nominal OC diameter & \multicolumn{2}{c|}{240 mm} \\ \hline
Nominal IC diameter & \multicolumn{2}{c|}{105 mm} \\ \hline
Nominal height ($\lambda/4$) & 931 mm & 466 mm \\ \hline
Active length & 210 mm & 190 mm \\ \hline
Aperture & \multicolumn{2}{c|}{30 mm} \\ \hline
\end{tabular}
\end{center}
\end{table}

\section{Fabrication of Prototypes}

\begin{figure}[tb]
\begin{center}
\setlength{\unitlength}{0.001in}
\begin{picture}(3065,2450)
\put(0,0){\includegraphics{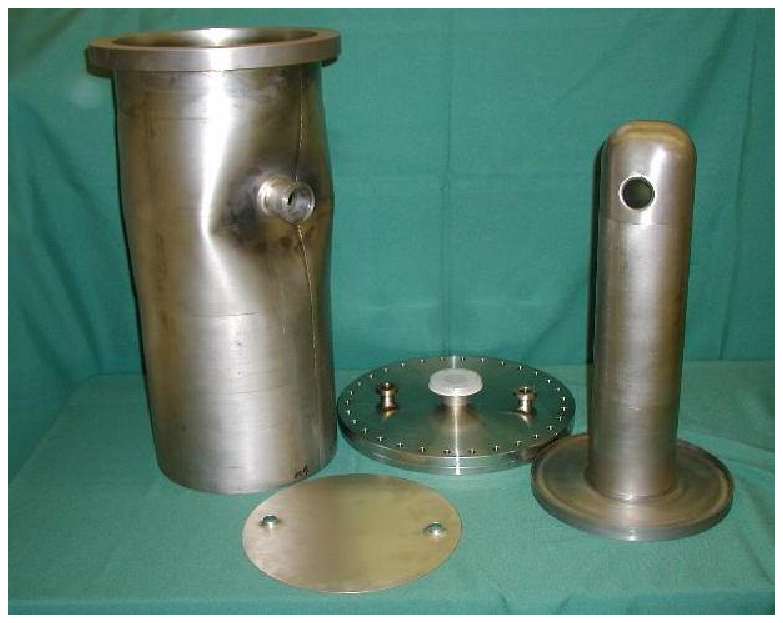}}
\put(2705,2140){\colorbox{white}{\makebox(300,250){\Large(a)}}}
\end{picture}\\[0.5ex]

\setlength{\unitlength}{0.001in}
\begin{picture}(3065,2560)
\put(0,0){\includegraphics{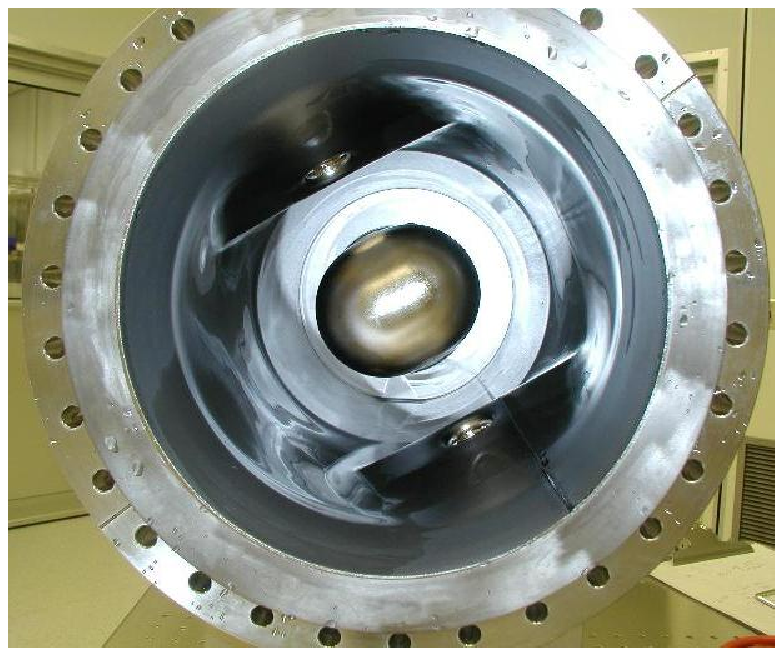}}
\put(0,2250){\colorbox{white}{\makebox(300,250){\Large(b)}}}
\end{picture}\\
\end{center}

\caption{(a) Nb parts for $\beta_m = 0.16$ QWR and (b) inside view of the
completed cavity.\label{F:pip}}
\end{figure}

Sheet Nb of thickness 2 mm and RRR $\geq 150$ was used.  The top plate, the
beam tubes, and the tip of the center conductor were machined from solid Nb. 
Holes were machined into the latter (see \autoref{F:dwg}) to improve the
contact with the liquid helium.  The bottom flange consists of a Nb-Ti ring
welded to the Nb outer conductor, mating with a stainless steel (SS) blank-off
flange; a Nb tuning plate is bolted to the outer conductor via this flange.  
Forming of the Nb parts was done at MSU and in the local area. Electron beam
welding was done by industry.  \autoref{F:pip} shows the Nb parts before and
after welding of the $\beta_m = 0.16$ cavity.  Indium joints were used to
provide a vacuum seal on the bottom flange and beam tube flanges.  Electrical
contact between the tuning plate and the outer conductor was made via pressure
from the bottom flange (thus the indium provides a vacuum seal, but not an RF
seal).  At $E_p = 20$ MV/m, the magnetic field at the joint is about 0.7 mT for
the $\beta_m = 0.16$ QWR and about 0.5 mT for the $\beta_m = 0.085$ QWR.

\begin{figure}[tb]
\incgraphLANDHALF{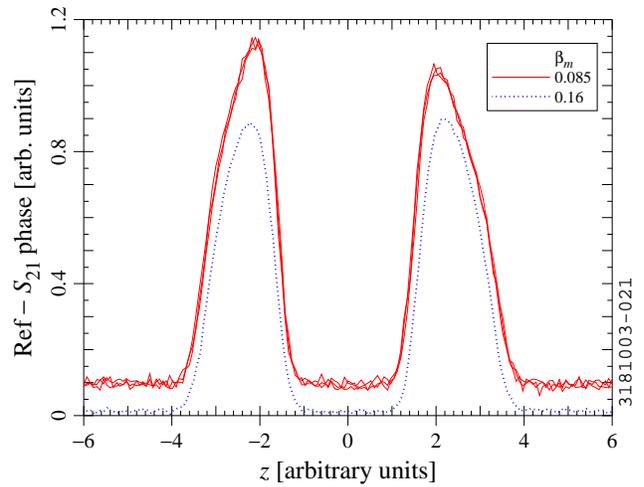}{1.0}

\caption{Bead pulls for the $\beta_m = 0.16$ and $\beta_m = 0.085$
QWRs.\label{F:bead}}
\end{figure}

Bead pulls were done to check the field flatness.  \autoref{F:bead} shows
the bead pull traces.  The field unflatness parameter ($\Delta E/E$) was 3.8\%
for the $\beta_m = 0.085$ cavity and 0.6\% for the $\beta_m = 0.16$ cavity.

\begin{figure}[tb]
\begin{center}
\setlength{\unitlength}{0.01in}
\begin{picture}(142,427)
\put(0,0){\includegraphics{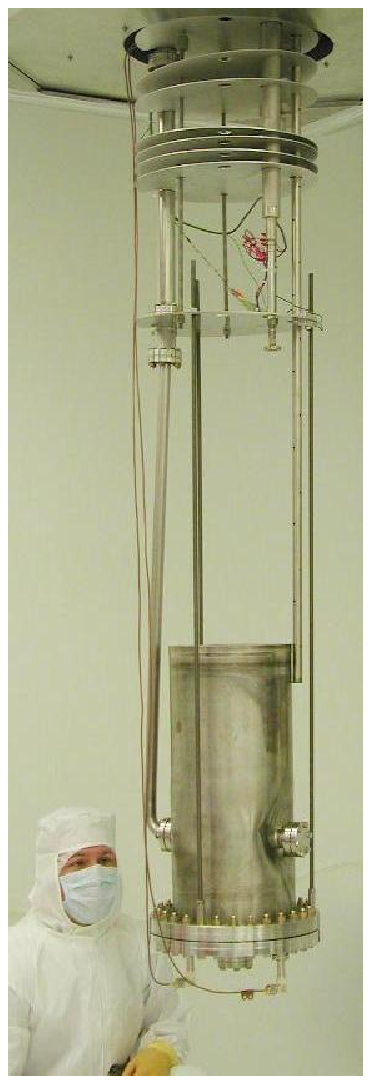}}
\put(0,399){\colorbox{white}{\makebox(30,25){\Large(a)}}}
\end{picture}
\setlength{\unitlength}{0.01in}
\begin{picture}(143,427)
\put(0,0){\includegraphics{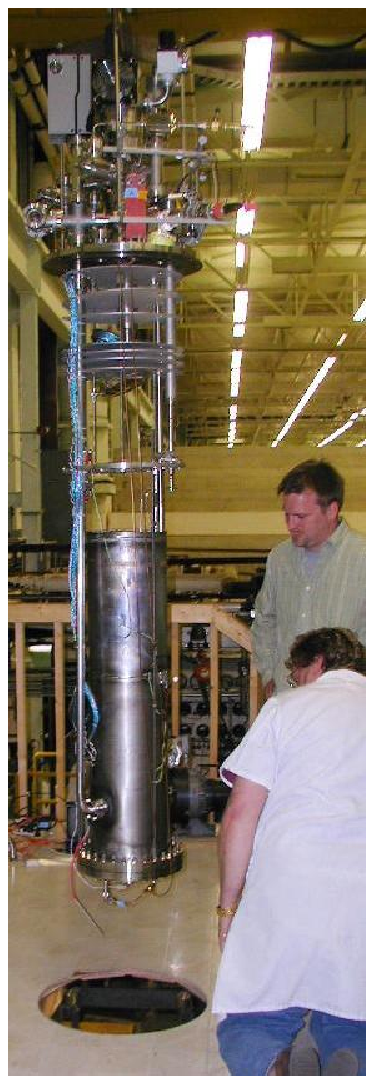}}
\put(107,399){\colorbox{white}{\makebox(30,25){\Large(b)}}}
\end{picture}
\end{center}

\caption{Photographs of (a) $\beta_m = 0.16$ QWR and (b) $\beta_m = 0.085$ QWR
on the RF test stand.\label{F:insert}}
\end{figure}

The completed QWRs were etched with a Buffered Chemical Polishing solution
(1:1:2 mixture by volume of concentrated hydrofluoric, nitric, and phosphoric
acid) to remove 120 $\mu$m from the inside surface.  The acid was circulated
through a chiller in a closed loop system to maintain a temperature $\leq
15$\degree C.  After etching, a high-pressure rinse with ultra-pure water was
done in a Class 100 clean room for 60 to 120 minutes.  The cavity was then
assembled onto an insert for subsequent RF testing.  \autoref{F:insert}a
shows the $\beta_m = 0.16$ cavity during assembly onto the insert. 
\autoref{F:insert}b shows the $\beta_m = 0.085$ cavity just prior to
insertion into the cryostat.

\section{RF Tests}

RF testing was done with the cavity immersed in a liquid helium bath at 4.2 K;
some additional measurements were done at 1.5 K or 2 K\@.  A phase feedback
loop was used to lock onto the resonance.  The RF power was provided by a 50 W
amplifier protected by a circulator.  Copper probe antennae (mounted on the
bottom flange, see \autoref{F:dwg}) were used to couple the power into the
cavity and pick up the transmitted power signal.  The input antenna length was
chosen to be near unity coupling at low field at 4.2 K.

Multipacting barriers were encountered at low field in both cavities.  We were
able to get through the barriers with 1 to 2 days of RF conditioning at 4.2
K\@.  The barriers were not completely eliminated by conditioning; 
reconditioning was required in some circumstances.  No conditioning was done at
higher temperatures.

\subsection{\texorpdfstring{$\beta_m = 0.16$}{beta(m) = 0.16} QWR}

\begin{figure}[b]

\begin{center}
\incgraphLANDHALFlab{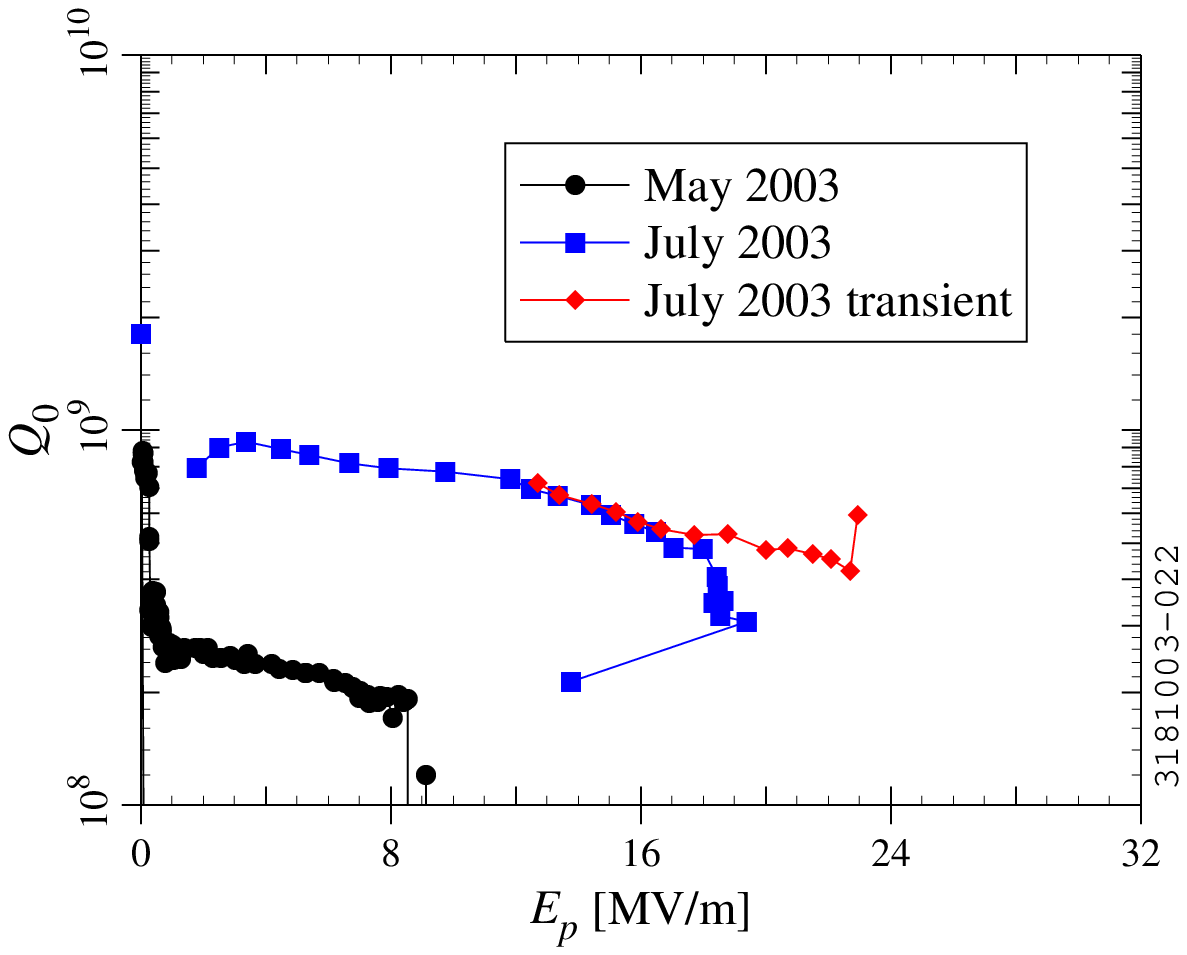}{1.0}{(a)}\\
\incgraphLANDHALFlab{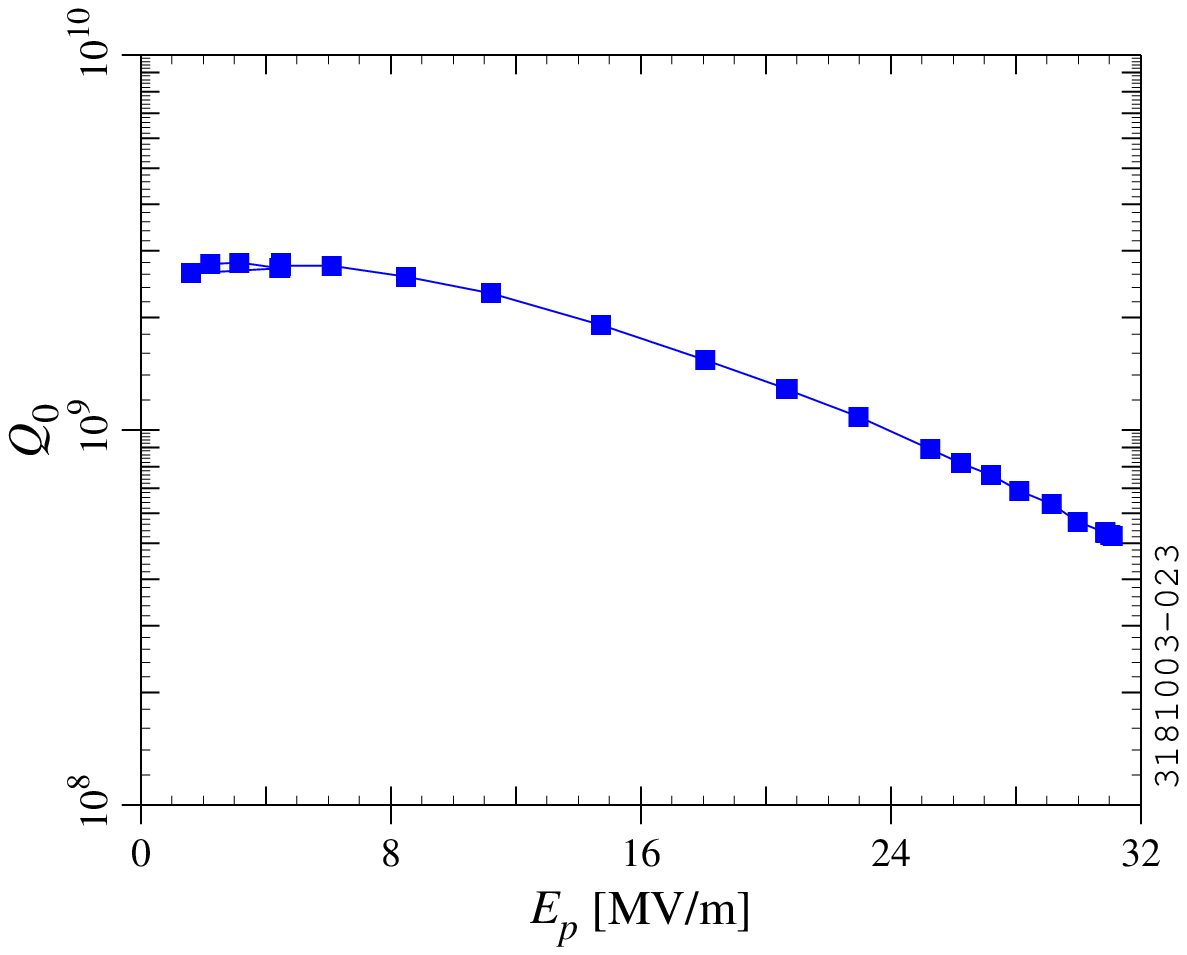}{1.0}{(b)}\\
\end{center}

\caption{RF tests of (a) the $\beta_m = 0.16$ QWR at 4.2 to 4.3 K and (b) the
$\beta_m = 0.085$ QWR at 4.2 K.\label{F:QvE}}
\end{figure}

The first RF test on the $\beta_m = 0.16$ QWR was done in May 2003.  Results are
shown in \autoref{F:QvE}a (circles).  The maximum field level reached in the
first test was $E_p = 8$ MV/m.  After the first test, the cavity was
disassembled, and grinding was done to remove a suspicious area on the shorting
plate.  An imbedded bit of foreign metal was also found and ground out.  After
grinding, the cavity was etched and rinsed again.

A second RF test was done in July 2003.  A field level of $E_p = 19$ MV/m was
reached in steady state (see \autoref{F:QvE}a, squares).  A higher field
($E_p = 23$ MV/m) could be reached with modulated RF (see \autoref{F:QvE}a,
diamonds).  These levels were reached after some improvement from RF
processing.  Helium processing was also attempted, but this did not further
improve the performance.  Modest x-ray signals were observed during the test
(110 mR/hour was the largest radiation level measured inside the radiation
shield).

A system of mirrors was used to view the outside of the cavity with a video
camera placed on top of the cryostat.  At high field, bubbles in the helium
bath were observed to come from the bottom flange of the cavity, indicating
that the dominant losses were not in the high magnetic field region.

As indicated in \autoref{F:QvE}a, there was a decrease in $Q_0$ at very low
field, possibly due to resistive losses in the RF joint between the outer
conductor and the tuning plate.  The low-field $Q_0$ value of $1.8 \cdot 10^9$
at 4.2 K corresponds to a surface resistance ($R_s$) of 19 n$\Omega$; the
expected contribution from the BCS term is 12 n$\Omega$.  The low-field $Q_0$
was $3 \cdot 10^9$ at $T = 2$ K, corresponding to a residual surface resistance
($R_0$) of 12 n$\Omega$.  Thus, the measured $R_s$ at 4.2 K is a bit smaller
than expected from the sum of the expected BCS $R_s$ and measured $R_0$.  For
$E_p > 2$ MV/m, the $Q_0$ was the same at 2 K as at 4.2 K.

\subsection{\texorpdfstring{$\beta_m = 0.085$}{beta(m) = 0.085} QWR}

Two changes were made for the $\beta_m = 0.085$ QWR test (in light of the 
problems encountered in the tests on the $\beta_m = 0.16$ QWR and the suspected
causes).  First, a ridge was added to the Nb tuning plate for better RF contact
with the outer conductor.  Second, a hollow tube was installed on the bottom
flange to touch the center of the tuning plate for improved heat sinking to the
helium bath.

The  first RF test on the $\beta_m = 0.085$ QWR was done in September 2003. 
Results are shown in \autoref{F:QvE}b.  A field level of $E_p = 31$ MV/m was
reached.  Above that field level, we reproducibly lost lock on the feedback
loop, possibly due to thermal breakdown.  The measured x-ray signals were $\leq
50$ mR/hour.  The video images indicated that bubbles were being nucleated more
or less uniformly near the top of the cavity, as one would expect for losses
due to the magnetic field.

The low-field $Q_0$ value of about $3 \cdot 10^9$ at 4.2 K corresponds to $R_s
= 6.3$ n$\Omega$; the expected contribution from the BCS term is 2.9 n$\Omega$.
The low-field $Q_0$ was $6 \cdot 10^9$ at $T = 1.5$ K, corresponding to $R_0 =
3.2$ n$\Omega$, in good agreement with the measured $R_s$ at 4.2 K and
expected BCS contribution.

\section{Conclusion}

A $\beta_m = 0.16$ QWR and a $\beta_m = 0.085$ QWR have been fabricated and
tested.  RF test results for the $\beta_m = 0.16$ QWR are marginal.  Further
tests with improved RF contact and better heat sinking of the Nb tuning plate
are planned. RF test results for the $\beta_m = 0.085$ QWR have exceeded the
design field level by a comfortable margin, with $Q_0 > 10^9$ at the design
field (at $T = 4.2$ K).  We were able to condition through low-field
multipacting barriers without a variable coupler (1 to 2 days of conditioning
were required).  The next steps in this QWR prototyping effort will be the
fabrication of complete cavities with integrated He vessels and the
characterisation of microphonics.

\section{Acknowledgements}

\small

Many thanks to all of the people at MSU and INFN-Legnaro who helped with
quarter-wave cavity design, prototyping, and testing.  A. Aizaz, J. Brandon, M.
Johnson, Sh.\ Alfredson Jones, Ch.\ Kolarchick Bieber, H. Laumer, D. Lawton, A.
McCartney, A. Moblo, S. Musser, D. Pedtke, J. Popielarski, J. Vincent, and R.
Zink provided valuable support for the project.  V. Andreev provided helpful
suggestions about RF joints.  Guidance from P. Kneisel and others at Jefferson
Lab has been extremely valuable in the development of the SRF program at MSU.

\end{document}